# UNIX RESOURCE MANAGERS

## Capacity Planning and Performance Issues[†]

Neil J. Gunther
*Performance Dynamics Consulting™*
*Castro Valley, California USA*
http://www.perfdynamics.com/

**ABSTRACT**
The latest generation of UNIX includes facilities for automatically managing system resources such as processors, disks, and memory. Traditional UNIX time-share schedulers attempt to be fair to all users by employing a round-robin policy for allocating CPU time. Unfortunately, a loophole exists whereby the scheduler can be biased in favor of a greedy user running many short CPU-demand processes. This loophole is not a defect but an intrinsic property of the round-robin scheduler that ensures responsiveness to the brief CPU demands associated with multiple interactive users. The new generation of UNIX constrains the scheduler to be equitable to all users regardless of the number of processes each may be running. This "fair-share" scheduling draws on the concept of pro rating resource "shares" across users and groups and then dynamically adjusting CPU usage to meet those share proportions. This simple administrative notion of statically allocating shares, however, belies the potential consequences for performance as measured by user response time and service level targets. We examine several simple share allocation scenarios and the surprising performance consequences with a view to establishing some guidelines for allocating shares.

**INTRODUCTION**

Capacity planning is becoming an increasingly important consideration for UNIX system administrators as multiple workloads are consolidated onto single large servers, and the procurement of those servers takes on the exponential growth of internet-based business.

The latest implementations of commercial UNIX to offer mainframe-style capacity management on enterprise servers include: AIX® Workload Manager (WLM), HP-UX® Process Resource Manager (PRM), Solaris® Resource Manager (SRM), SGI and Compaq. The ability to manage server capacity is achieved by making significant modifications to the standard UNIX operating system so that processes are inherently tied to specific users. Those users, in turn, are granted only a certain fraction of system resources. Resource usage is monitored and compared with each user's grant to ensure that the assigned entitlement constraints are met.

Shared system resources that can be managed in this way include: processors, memory, and mass storage. Prima face, this appears to be exactly what is needed for a UNIX system administrator to manage server capacity. State of the art UNIX resource management, however, is only equivalent to that which was provided on mainframes more than a decade ago but it had to start somewhere. The respective UNIX operations

---

manuals spell out the virtues of each vendor's implementation, so the purpose of this paper is to apprise the sysadm of their limitations for meeting certain performance requirements.

For example, an important resource management option is the ability to set entitlement *caps* when actual resource consumption is tied to other criteria such as performance targets and financial budgets. Another requirement is the ability to manage capacity across multiple systems e.g., shared-disk and NUMA clusters. As UNIX system administrators become more familiar with the new automated resource management facilities, it also becomes important that they learn some of the relevant capacity planning methodologies that have already been established for mainframes.

In that spirit, we begin by clearing up some of the confusion that has surrounded the motivation and the terminology behind the new technology. The common theme across each of the commercial implementations is the introduction of a *fair-share* scheduler. After reviewing the alternative scheduler, we examine some of the less obvious potential performance pitfalls. Finally, we present some capacity planning guidelines for migrating to automated UNIX resource management.

**LOOPHOLES IN TIME**

UNIX is fundamentally a time-share (TS) operating system [Ritchie & Thompson 1978] aimed at providing responsive service to multiple interactive users. Put simply, the goal of generic UNIX is to create the illusion for each user that they are the only one accessing system resources. This illusion is controlled by the UNIX time-share scheduler.

A time-share scheduler attempts to give all active processes equal access to processing resources by employing a *round-robin* scheduling policy for allocating processor time. Put simply, each process is placed in a run-queue and when it is serviced, it gets a fixed amount of time running on the processor. This fixed time is a called a service *quantum* (commonly set to 10 milliseconds).

Processes that demand less than the service quantum, generally complete processor service without being interrupted. However, a process that exceeds its service quantum has its processing interrupted and is returned toward the back of the run-queue to await further processing. A natural consequence of this scheduling policy is that processes with shorter processing demands are favored over longer ones because the longer demands are always preempted when their service quantum expires.

Herein lies a loophole. The UNIX scheduler is biased intrinsically in favor of a greedy user who runs many short-demand processes. This loophole is not a defect, rather it is a property of the round-robin scheduler that implements time-sharing which, in turn, guarantees responsiveness to the short processing demands associated with multiple interactive users. Figure 1 shows this effect explicitly.

In Fig. 1, two users are distinguished as owning *heavy* and *light* processes in terms of their respective processing demands. Below 10 processes, the heavy user sees a better transaction throughput than the light user. Beyond 10 processes the throughput of the light user begins to dominate the heavy user; in spite of the fact that both users are adding more processes into the system!

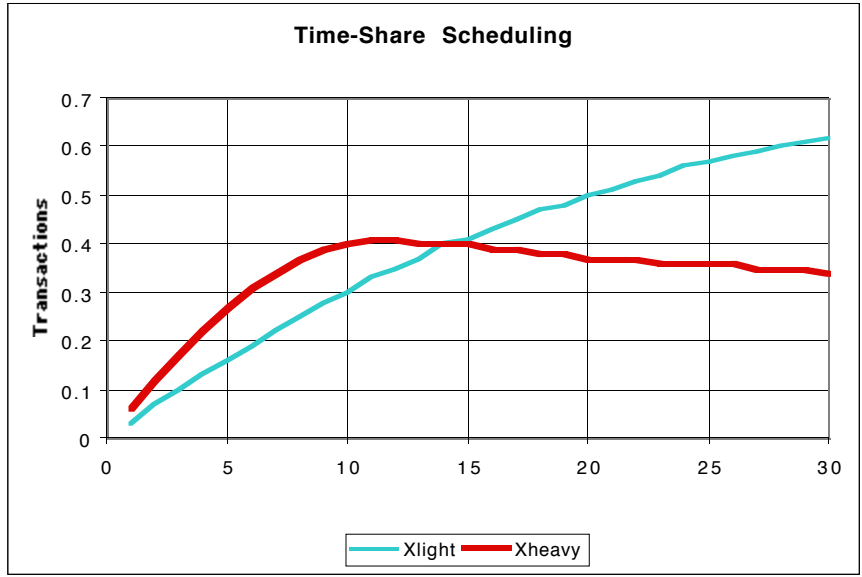

Fig. 1. Generic UNIX intrinsically favors shorter service-demand processes.

Note that the UNIX scheduler schedules processes, not users. Under a time-share scheduler there is no way for users to assert that they would like a more equitable allocation of processing resources to defeat greedy consumption by other users. The new generation of UNIX addresses this loophole by implementing a *fair-share* (FS) scheduler [Kay & Lauder 1988] that equitable processing to all *users* regardless of the number of processes each may be running.

**WHAT IS FAIR?**

To define "fairness", the FS-based scheduler has a control parameter that allows the sysadm to pro rate literal "shares" across users and groups of users. The FS scheduler then dynamically adjusts processor usage to match those share proportions. Similarly, shares can also be allocated to control memory and disk I/O resource consumption. For the remainder of this paper, we'll confine the discussion to processor consumption because it is easiest to understand and has the most dynamic consequences.

In the context of computer resources, the word *fair* is meant to imply *equity*, not *equality* in resource consumption. In other words, the FS scheduler is *equitable* but not *egalitarian*. This distinction becomes clearer if we consider shares in a publicly owned corporation. Executives usually hold more *equity* in a company than other employees. Accordingly, the executive shareholders are entitled to a greater percentage of company profits. And so it is under the FS scheduler. The more equity you hold in terms of the number of allocated resource shares, the greater the percentage of server resources you are entitled to at runtime.

The actual number of FS shares you own is statically allocated by the UNIX sysadm. This share allocation scheme seems straightforward. However, under the FS scheduler there are some significant differences from the way corporate shares work. Although the shares allocated to you is a *fixed* number, the proportion of resources you receive may vary *dynamically* because your entitlement is calculated as a function of the total number of *active* shares; not the total pool of shares.

**CAPPING IT ALL OFF**

Suppose you are entitled to receive 10% of processing resources by virtue of being allocated 10 out of a possible 100 system-wide processor shares. In other words, your processing *entitlement* would be 10%. Further suppose you are the only user on the system. Should you be entitled to access 100% of the processing resources? Most UNIX system adminstrators believe it makes sense (and is fare) to use of all the resources rather than have a 90% idle server.

But how can you access 100% of the processing resources if you only have 10 shares? You can if the FS scheduler only uses *active* shares to calculate your entitlement. As the only user active on the system, owning 10 shares out of 10 active shares is tantamount to a 100% processing entitlement.

This natural inclination to make use of otherwise idle processing resources really rests on two assumptions:

1. You are not being **charged** for the consumption of processing resources. If your manager only has a budget to pay for a maximum of 10% processing on a shared server, then it would be fiscally *undesirable* to exceed that limit.
2. You are unconcerned about **service targets**. It's a law of nature that users complain about perceived changes in response time. If there are operational periods where response time is significantly better than at other times, those periods will define the future service target.

One logical consequence of such dynamic resource allocation, is the inevitable loss of control over resource consumption altogether. In which case, the sysadm may as well spare the effort of migrating to an FS scheduler in the first place. If the enterprise is one where chargeback and service targets are important then your entitlement may need to be clamped at 10%. This is achieved through an additional *capping* parameter. Not all FS implementations offer this control parameter.

The constraints of chargeback and service level objectives have not been a part of traditional UNIX system administration, but are becoming more important with the advent of application consolidation, and the administration of large-scale server configurations (e.g., like those used in major e-commerce websites). The capping option can be very important for enterprise UNIX capacity planning.

**CONSPICUOUS CONSUMPTION**

Let's revisit the earlier example of a heavy user and light user, but this time running their processor-bound jobs under an FS scheduler with the heavy user awarded a 90% processor allocation and the light user allocated only 10% of the processor. Contrary to the TS scheduling effect, the light user should not be able to dominate the heavy user by running more short-demand processes.

In fact, Figure 2 bares this out. The heavy user continues to maintain a higher transaction throughput because they have a 90% allocation of cpu resources. Even as the number of processes is dramatically increased, the light user cannot dominate the heavy user because they are only receiving 10% of cpu resources.

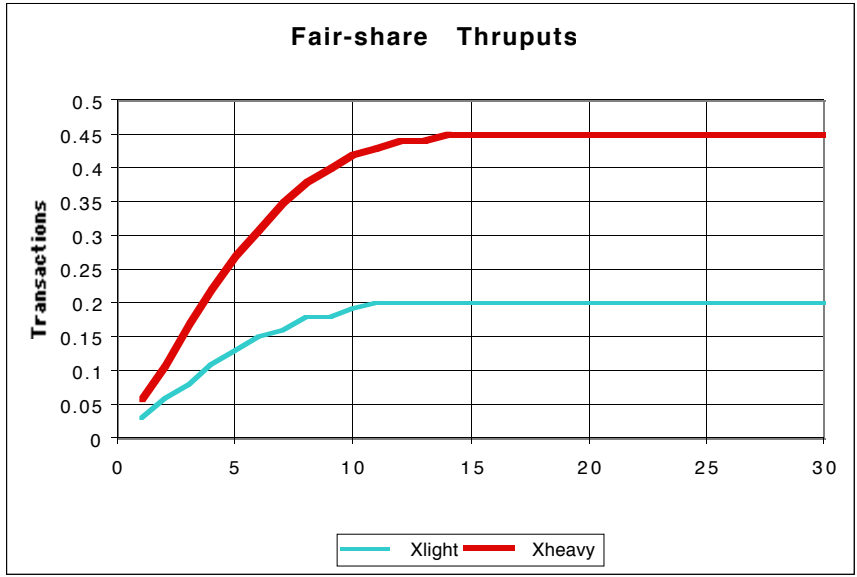

Fig. 2.   The FS scheduler does not favor shorter CPU-demand processes.

The effect is easier to interpret if we look at comparative utilization. Under the FS scheduler the utilization of each user climbs to match their respective cpu entitlements at about 10 processes.

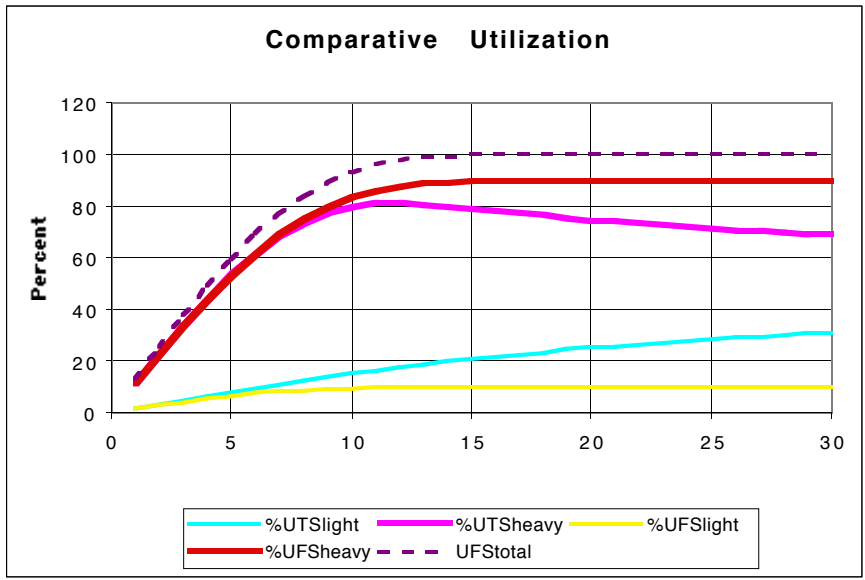

Fig. 3.   Comparative CPU utilization for two users with processes running under the time-share and fair-share schedulers.

Contrast this with a generic TS scheduler, and we see the heavy user reaches a maximum utilization of about 80% and then falls off as cpu capacity is usurped by the light user as it continues to consume cpu above 10%. Under either scheduling policy, the dashed line shows that total processor utilization reaches 100%, as expected.

**NICE GUYS FINISH LAST**

The reader may wondering why we haven't said anything about two well-known aspects of UNIX process control: priorities and nice values. Couldn't the same results be achieved under a TS scheduler together with a good choice of nice values? The answer is, no and here's why.

---

**Solaris® SRM Fair Share Algorithm**

**User Level Scheduling**

*Every 4000 milliseconds, accumulate the CPU usage per user*

```
Parameters:
   Delta = 4 Sec;
   DecayUsage;

   for(i=0; i<USERS; i++) {
        usage[i] *= decayUsage;
        usage[i] += cost[i];
        cost[i] = 0;
   }
```

**Process Level Scheduling**

```
Parameters:
   priDecay = Real number in the range [0..1]
```

*Every 1000 milliseconds, decay ALL process priority values*

```
   for(k=0; k<PROCS; k++) {
        sharepri[k] *= priDecay;
   }

   priDecay = a * p_nice[k] + b;
```

*Every 10 milliseconds (or tick) alter the active process priorities per user.*

```
   for(i=0; i<USERS; i++) {
        sharepri[i] += usage[i] * p_active[i];
   }
```

---

It's true that the TS scheduler has a sense of different priorities, but these priorities relate only to processes, and not users. The TS scheduler employs an internal priority scheme (there are several variants in generic UNIX. See Chapter 7 in Mc Dougall, et al. 1999 for a brief overview) to adjust the position of processes waiting in the run-queue so that it can meet the goal of creating the illusion for each user that they are the only one accessing system resources.

In standard UNIX, using the *nice* command to "renice" processes (by assigning a value in the range –19 to +19) merely acts as a *bias* for the TS scheduler when it repositions processes in the run-queue. This scanning and adjustment of the run-queue occurs about once every 10 milliseconds (or 1 processor tick). Once again, in standard UNIX, these internal priority adjustments apply only to processes and how well they are meeting the time-share goal. There is no association between processes and the users that own them. Hence, standard UNIX is susceptible to the loophole mentioned earlier in this paper.

The FS scheduler, in contrast, has two levels of scheduling: process and user. (See table) The process level scheduling is essentially the same as that in standard UNIX and nice values can be applied there as well. The user level scheduler is the major new component and the relationship between these two pieces is presented in the form of (highly simplified) pseudo-code in the following table (for Solaris® SRM).

Notice that whereas the process-level scheduling still occurs 100 times a second, the user-level scheduling adjustments (through the `cpu.usage` parameter) are made on a time-scale 400 times longer, or once every 4 seconds. In addition, once a second, the process-level priority adjustments that were made in the previous second begin to be "forgotten". This is necessary to avoid starving a process of further processor service simply because it exceeded its share in some prior service period [Bettison et al. 1991].

**PERFORMANCE PITFALLS**

So far, we've outlined how the FS scheduler provides a key mechanism for capacity management of processing resources. What about performance? Having the ability to allocate resources under a scheduler that can match this allocation to actual resource consumption still tells us nothing about the impact on user response times.

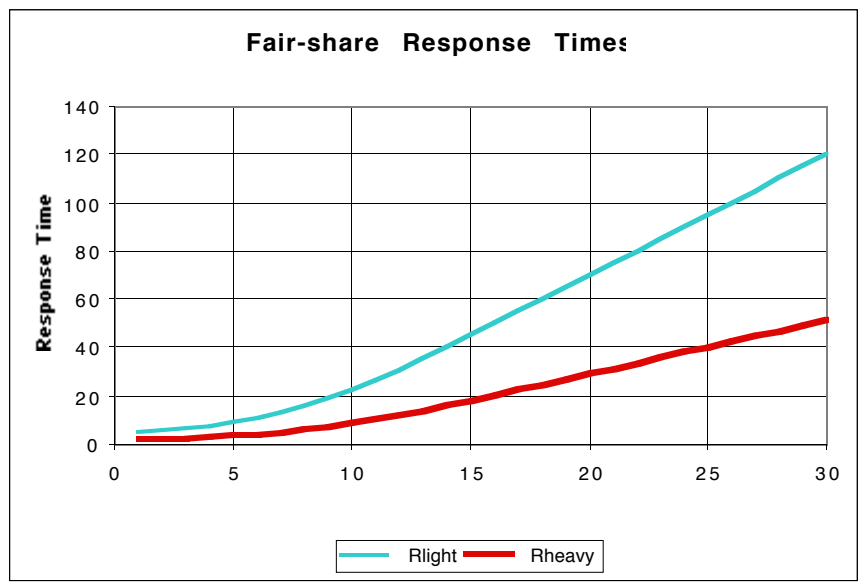

Fig. 5. Response times of heavy and light users under fair-share.

Moreover, none of the current FS implementations seriously discuss the performance impact of various parameter settings, yet this knowledge is vital for the management of many commercial applications. To underscore the importance of this point, compare the difference in response times between the TS scheduler in Figure 4 and the FS scheduler shown in Figure 5.

Under TS we see that the light user always has the better response time than the heavy user, while under FS where the light user only has a 10% cpu entitlement, the opposite is true. This is correct FS behavior but consider the impact on the light user if they were migrated off a generic UNIX server to a consolidated server running the FS scheduler. That user would only get 10% of the processor and their corresponding response time would be degraded by a factor of six times. That would be like suddenly going from a 56Kbps modem to a 9600 baud modem!

**PERFORMANCE UNDER STRESS**

Next, we consider the performance impact on multiple users from groups of users with larger CPU entitlements. Suppose there are three groups: a database group (DBMS), a web server group (WEB), and three individual users: usrA, usrB and usrC in a third group.

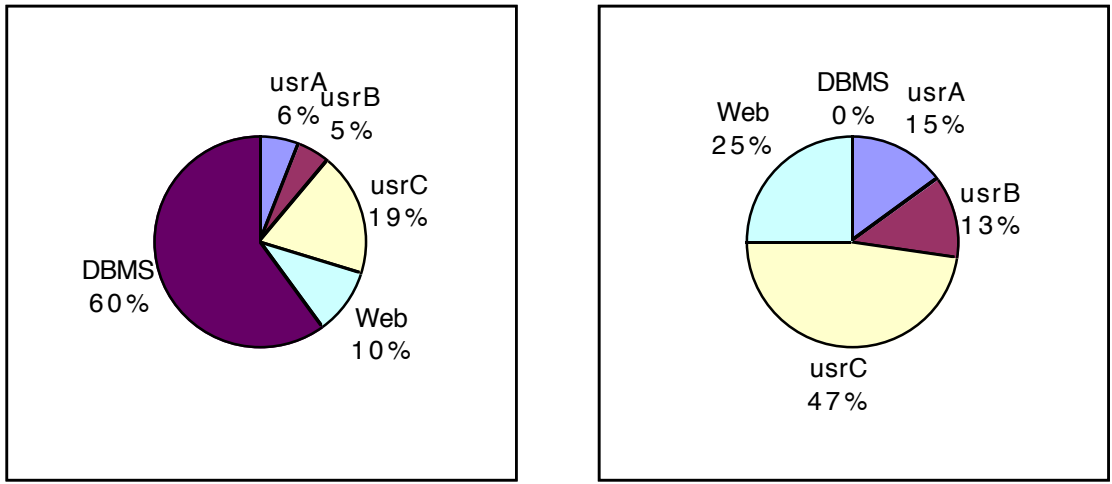

Fig. 6a and 6b. Entitlements for DBMS=60, Web=10, Users=30 shares.

The group allocations of the 100 pool shares is: DBMS = 60 shares, Web = 10 shares, Users = 30 shares. The corresponding entitlements are shown in Figure 6 with the users A,B,C awarded 6, 5, and 19 cpu.shares respectively

If all the users and groups are **active** on the system and running processor-bound workloads (a worst case scenario), the processor utilization exactly matches the entitlements of Figure 6a.

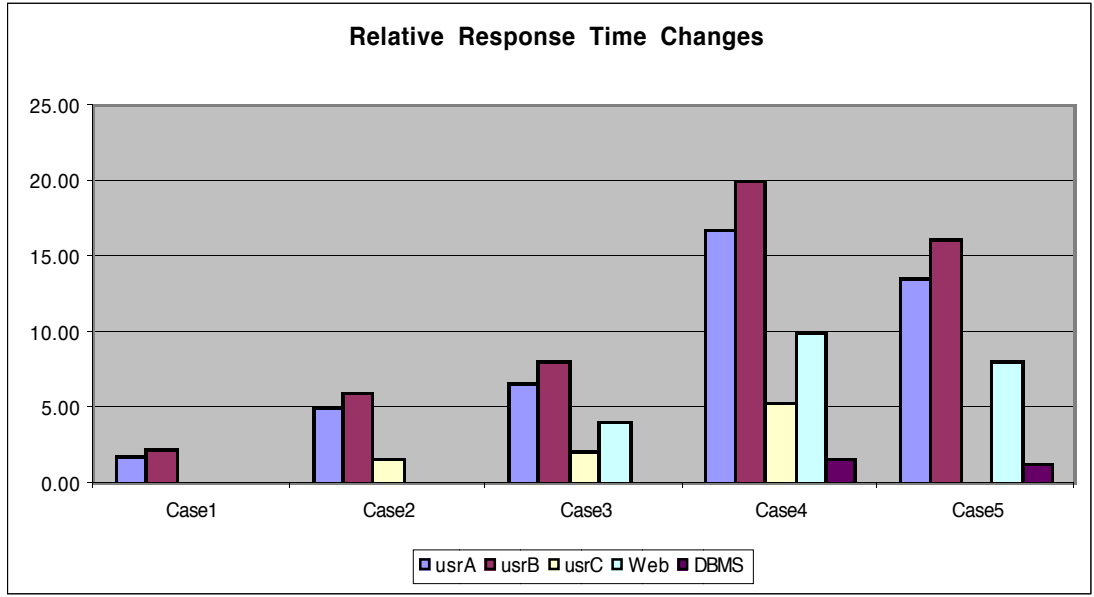

Fig. 7. Relative impact on users A and B as load is added to the system.

If the DBMS group becomes **inactive**, then without capping, the entitlements would be dynamically adjusted to those shown in Figure 6b. These new fractions are not intuitively obvious even for CPU-bound workloads and could present a surprise to a sysadm.

More surprising still is what happens to response times. The relative response times cannot be calculated purely from the static share or entitlement allocations and must be derived from a performance model. Figure 7 shows the effect on the average response times seen by users A and B as other users and groups (see Fig. 6) become active on the system. We see immediately that by the time all groups become active, the response times of A and B have increased (degraded) by a factor of 10x.

This is a worst-case scenario because it assumes that all work is CPU-bound. But worst-case analysis is precisely what needs to be done in the initial stages of any capacity planning exercise.

All of the capacity scenarios presented in this paper were created using PDQ© [Gunther 1998] and have been corroborated with TeamQuest Model [TeamQuest 1998]. In general, none of the current UNIX resource management products offer a way to look ahead at the performance impact of various share allocations. More sophisticated tools like PDQ and Model are needed.

**MIGRATION TIPS and TOOLS**

It should be clear by now that to bridge the gap between allocating capacity and the respective potential performance consequences, some kind of performance analysis capability is important for automated resource management. As the next table shows, none of the top vendors provide such tools.

What size share pool should be used? The total pool of shares to be allocated on a server is an arbitrary number. One might consider assigning a pool of 10,000 shares so as to accurately express entitlements like 1/3 as a decimal percentage with two significant digits i.e., 33.33%. It turns out that the FS scheduler is only accurate to within a few percent. Since it's easier to work with simple percentages, a good rule of thumb is to set the total pool of shares to be 100. In the above example, an entitlement of one third would be simply be equal to 33%.

| *O/S* | *Manager* | *Tools* |
|---|---|---|
| **AIX** | WLM | SMIT |
|  |  | RSI |
| **HP-UX** | PRM | prmanalyze |
|  |  | Glance$^®$ |
| **Solaris** | SRM | srmstat |
|  |  | liminfo |
|  |  | limadm |
|  |  | limreport |

When contemplating the move to enterprise UNIX resource management, you will need to reflect how you are going to set the FS parameters. These vary by product but some general guidelines can be elucidated. In particular:
- ❖ Start collecting appropriate consumption measurements on your current TS system. Collect process data to measure CPU and memory usage. This can be translated directly into a first estimate for respective entitlements.
- ❖ Allocate shares or entitlements to your "loved ones" (mainframe-speak) or your most favored users, first.
- ❖ Allocate shares based on the *maximum* response times seen under TS. This approach should provide some wiggle room to meet service targets.
- ❖ Turn the capping option on (if available) when service targets are of primary importance.
- ❖ If service targets are not being met after going over to an FS scheduler, move the least favored users and resource groups onto another server or invoke processor domain (if available).

Use modeling tools such as PDQ© [PDQ 1999] and TeamQuest Model® to make more quantitative determinations of how each combination of likely share allocations pushes on the performance envelope of the server.

**COMMERCIAL COMPARISONS**

The following table summarizes the key feature differences between the three leading commercial UNIX resource managers.

MVS has been included at the end of the table for the sake of completeness. All of the commercial UNIX resource managers are implemented as fair-share schedulers. IBM/AIX, Compaq and HP developed their own implementations while SGI and Sun use a modified variant of the SHARE II product [Bettison et al. 1991] licensed from Aurema Pty Ltd. Neither SGI SHARE-II nor Solaris SRM offer resource capping [Gunther 1999].

Although each product has tools that are intended to help administer resource allocation, they tend to be somewhat ad hoc rather than tightly integrated with the operating system. This should improve with later releases.

| O/S | Manager | FS | Capping | Parameters |
|---|---|---|---|---|
| **AIX** | WLM | Yes | Yes | Targets/Limits |
| **Compaq** | Tru64 | Yes | No | Shares |
| **HP-UX** | PRM | Yes | Yes | Entitlements |
| **IRIX** | SHARE II | Yes | No | Shares |
| **Solaris** | SRM | Yes | No | Shares |
| **MVS** | WLM-CM | Yes | Yes | Service classes |
| **MVS** | WLM-GM | No | N/A | Service goals |

Both AIX [Brown 1999] and Solaris [McDougall et al. 1999] tend to present their management guidelines using the terminology of *shares*. But shares are really the underlying mechanism by which usage comparisons are made. It would be preferable to discuss system administration in terms of entitlements (which are percentages) because it is easier for a sysadm to match entitlement percentages against the measured utilization percentages of the respective subsystems.

MVS Workload Manager in Compatibility Mode (WLM–CM) is the vestige of the original OS/390 System Resource Manager (also called SRM) first implemented a decade ago [Samson 1997] by IBM. It is this level of functionality that the enterprise UNIX vendors have introduced. The newer and more complex MVS in Goal Mode (WLM–GM) was only introduced by IBM a few years ago and it is likely to be several more years before automated UNIX resource managers reach that level of maturity.

**SUMMARY**

Commercial UNIX is starting to emulate the kind of resource management facilities that have become commonplace on mainframes. Some of the latest UNIX implementations to offer capacity management on enterprise servers include: AIX® Workload Manager (WLM), HP-UX® Process Resource Manager (PRM), and Solaris® Resource Manager (SRM). Nevertheless, the UNIX technology is still a decade behind that found on current mainframes; a key distinction being the ability to define specific service goals for mainframe workloads.

The latter is accomplished by having the scheduling mechanism monitor the wait-time for each service group in the run-qeueue, whereas, the FS scheduler only monitors the busy-time (CPU usage) consumed when each user or group is in service. This limitation makes it difficult to allocate resources in such a way to meet service targets; an important requirement for enterprise UNIX running commercial applications.

Until service goal specifications are engineered into UNIX, the sysadm must pay careful attention to the allocation of shares and entitlements. Better tools of the type discussed in this paper, would go a long way toward making the sysadm's job easier.